\documentclass[longbibliography,physrev,aps,twocolumn]{revtex4-2}
\usepackage{graphicx,epsfig}
\usepackage{times}
\usepackage{amsmath,amssymb,wasysym}
\usepackage{dsfont} 
\usepackage{color}
\usepackage[linktocpage=true,colorlinks=true, pdfborder={0 0 0},linkcolor=blue,citecolor=blue,urlcolor=blue,anchorcolor=black,linkcolor=blue]{hyperref}

\newcommand{\op}[1]{\hat{#1}}

\newcommand{\normV}[1]{||{#1}||_2}
\newcommand{\normM}[1]{||{#1}||}
\newcommand{\normspec}[1]{||{#1}||_{\text{2}}}

\newcommand{\ident}{\mathds{1}}
\newcommand{\C}{\mathbb{C}}

\newcommand{\quoting}[1]{``#1''}

\newcommand{\rem}[1]{}

\newcommand{\imagc}[1]{\text{Im}\,#1}

\newcommand{\bra}[1]{\langle#1|}
\newcommand{\ket}[1]{|#1\rangle}

\newcommand{\braket}[2]{\langle#1|#2\rangle}

\newcommand{\normFro}[1]{||#1||_{\text{F}}}
\newcommand{\rca}{\xi}

\newcommand{\pseudospectrum}{\Lambda}

\DeclareMathOperator{\trace}{Tr}

\newcommand{\Hperturbed}{\op{H}}
\newcommand{\Hs}{\op{H}_0}
\newcommand{\Hp}{\op{H}_1}

\newcommand{\order}{n}

\newcommand{\ev}{E}
\newcommand{\evn}{E^{(0)}}
\newcommand{\evEP}{\ev_{\text{EP}}}
\newcommand{\evEPk}[1]{\ev_{\text{EP},{#1}}}

\newcommand{\state}{R}

\newcommand{\GF}{\op{G}}

\newcommand{\pseps}{\varepsilon}

\newcommand{\ind}{l}

\newcommand{\PF}{K}
\newcommand{\HL}[1]{#1}
\newcommand{\rcanum}{\rca_{\text{num}}}
\newcommand{\lw}{\gamma}

\newcommand{\roneM}{\op{W}}
\newcommand{\Ea}{E_{\text{a}}}
\newcommand{\Eb}{E_{\text{b}}}
\newcommand{\freqisland}{\Omega_{\text{is}}}
\newcommand{\freqchaotic}{\Omega_{\text{ch}}}
\newcommand{\projector}{\op{P}}
\newcommand{\nilpotent}{\op{N}}

\newcommand{\crit}{c}

\begin{document}

\title{Moving along an exceptional surface towards a higher-order exceptional point}
\author{Jan Wiersig}
\affiliation{Institut f{\"u}r Physik, Otto-von-Guericke-Universit{\"a}t Magdeburg, Postfach 4120, D-39016 Magdeburg, Germany}
\email{jan.wiersig@ovgu.de}
\date{\today}
\begin{abstract}
Open systems with non-Hermitian degeneracies called exceptional points show a significantly enhanced response to perturbations in terms of large energy splittings induced by a small perturbation. This reaction can be quantified by the spectral response strength of the exceptional point. We extend the underlying theory to the general case where the dimension of the Hilbert space is larger than the order of the exceptional point. This generalization allows us to demonstrate an intriguing phenomenon: The spectral response strength of an exceptional point increases considerably and may even diverge to infinity under a parameter variation that eventually increases the order of the exceptional point. This dramatic behavior is in general not accompanied by a divergence of the energy eigenvalues and is shown to be related to the well-known divergence of Petermann factors near exceptional points. Finally, an accurate and robust numerical scheme for the computation of the spectral response strength based on the general theory and residue calculus is presented. 
\end{abstract}
\maketitle

\section{Introduction}
\label{sec:intro}
Exceptional points (EPs) are important and still not fully exploited non-Hermitian degeneracies in open quantum and wave systems. At an EP of order~$\order$ (EP$_\order$) exactly~$\order$ eigenenergies and the corresponding energy eigenstates  coalesce~\cite{Kato66,Heiss00,Berry04,Heiss04,MA19}. This is in contrast to a conventional degeneracy where only eigenenergies coalesce while the eigenstates can be chosen to be orthogonal. The existence of an EP requires not only the non-Hermiticity of the Hamiltonian, $\op{H} \neq \op{H}^\dagger$, but also nonnormality, i.e., $[\op{H},\op{H}^\dagger] \neq 0$.
EPs have been observed experimentally in diverse physical systems~\cite{DGH01,DDG04,DFM07,LYM09,POL14,POL16,RZZ19,CKL10,RBM12,GEB15,SKM16,WHL19}. 
Moreover, EPs have potential applications, in particular in optics and photonics~\cite{MA19}, such as loss-induced suppression of lasing~\cite{POR14}, mode discrimination in multimode lasing~\cite{HMH14}, orbital angular momentum microlasers~\cite{MZS16}, unidirectional lasing~\cite{POL16}, mode conversion~\cite{XMJ16,DMB16}, circularly-polarized light sources~\cite{RMS17}, chiral perfect absorption~\cite{SHR19}, optical amplifiers with improved gain-bandwidth product~\cite{ZOE20}, subwavelength control of light transport~\cite{XLJ23}, and as a resource for hardware encryption~\cite{YZZ23}.  

One particular relevant application is that of EP-based sensors~\cite{Wiersig14b,COZ17,HHW17,XLK19,LLS19,WLY20,Wiersig20b,Wiersig20c,KW21,KCE22}. These potentially ultrasensitive devices take advantage of the significant spectral response to generic perturbations: A system with an EP$_\order$ experiences an energy splitting proportional to the $\order$th root of the perturbation strength~$\varepsilon$~\cite{Kato66}. For sufficiently small~$\varepsilon$ this is larger than the linear scaling near a conventional degeneracy. The strength of the response to small perturbations can be characterized by a single quantity, the spectral response strength $\rca$~\cite{Wiersig22,Wiersig22b, Wiersig22c}. A large $\rca$ signals a significant spectral response to generic perturbations. 
The theory in Ref.~\cite{Wiersig22} is restricted to the special case where the order of the EP, $\order$, equals the dimension of the considered Hilbert space, $m$.

There are special, nongeneric perturbations that leave the given system on an EP$_\order$. In parameter space these perturbations generate a, possibly higher-dimensional, manifold, the so-called exceptional surface~\cite{ZRK19,ZNO19}. Such surfaces may remove the harmful consequences of fabrication intolerances and have been investigated for optical amplifiers~\cite{ZOE20}, sensing~\cite{QXZ21}, control of spontaneous emission~\cite{ZHO21}, and chiral perfect absorbers~\cite{SZM22}.

The spectral response of an isolated eigenvalue (not part of an EP) can be quantified by the Petermann factor~$\PF$~\cite{HS20}.  
It was originally introduced to measure the linewidth broadening resulting from quantum excess noise in lasers~\cite{Petermann79,Siegman89a,Siegman89b,Siegman95,LDM97,LEM98,Schomerus09}.
The Petermann factors near an EP are determined by its spectral response strength~$\rca$ and the energy splitting~\cite{Wiersig23,HS23}.  

\begin{figure}[ht]
	\includegraphics[width=0.95\columnwidth]{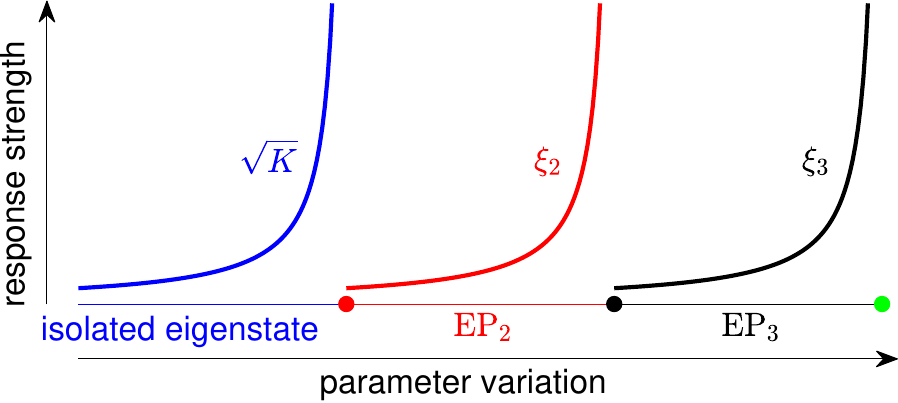}
	\caption{Illustration of the divergence of the spectral response strength of an isolated eigenvalue (Petermann factor $\PF$) and exceptional points of order~$\order$ (EP$_\order$ with spectral response strength~$\rca_\order$) when approaching a higher-order EP. The dots mark a transition from an isolated eigenvalue to an EP$_2$ and from an EP$_\order$ to an EP$_{\order+1}$.}
	\label{fig:kxi}
\end{figure}

The aim of this paper is twofold. First, we extend the theory of spectral response strength~$\rca$ to the general case $m \geq \order$.
Second, we use this generalized theory to demonstrate that $\rca$ may diverge to infinity when the system is moving along an exceptional surface towards an EP of higher order. 
We show that this is not only reminiscent to the behavior of the Petermann factor~$\PF$ when approaching an EP~\cite{Berry03} but that it is the same behavior originating from the fact that $\sqrt{\PF}$ itself can be seen as a spectral response strength of isolated eigenvalues. 
Figure~\ref{fig:kxi} illustrates these findings.

The paper is structured as follows. Section~\ref{sec:srsEP} reviews the spectral response strength as introduced in Ref.~\cite{Wiersig22}. Section~\ref{sec:general} generalize this definition to higher-dimensional Hilbert spaces and Sec.~\ref{sec:divergence} reveals the divergence of the spectral response strength of EPs that approach an EP of higher order. An instructive toy model is provided in Sec.~\ref{sec:toymodel} and a realistic photonic example in Sec.~\ref{sec:example}. In Sec.~\ref{sec:Petermann} the relation to the divergence of the Petermann factor is discussed. Based on the general theory and residue calculus a numerical scheme for the spectral response strength is developed in Sec.~\ref{sec:residue}. A conclusion is given in Sec.~\ref{sec:conclusion}.

\section{Spectral response strength}
\label{sec:srsEP}
Before turning to the general case, we briefly review the derivation of the spectral response strength associated to an $\order\times\order$ Hamiltonian at an EP$_\order$~\cite{Wiersig22}. Central are two matrix norms; see, e.g., Ref.~\cite{HJ13}. The first is the Frobenius norm 
\begin{equation}\label{eq:Fronorm}
	\normFro{\op{A}} := \sqrt{\trace{(\op{A}^\dagger\op{A})}}  = \sqrt{\sum_{ij}|A_{ij}|^2}
\end{equation}
where $\trace$ is the trace and $A_{ij}$ are the matrix elements of the linear operator~$\op{A}$ in any orthonormal basis. The second matrix norm is the spectral norm 
\begin{equation}\label{eq:defspn}
	\normspec{\op{A}} := \max_{\normV{\psi} = 1}\normV{\op{A}\psi} \ .
\end{equation}
We use the notation $\normspec{\cdot}$ both for the spectral norm of a matrix and the 2-norm $\normV{\psi} = \sqrt{\braket{\psi}{\psi}}$ of a vector~$\ket{\psi}$ based on the conventional inner product in complex vector space. 
Both matrix norms are compatible with the vector 2-norm, i.e.,
\begin{equation}\label{eq:compatible}
	\normV{\op{A}\psi} \leq \normM{\op{A}}\,\normV{\psi} 
\end{equation}
and both share the property of unitary invariance, i.e., 
\begin{equation}\label{eq:unitary}
	\normM{\op{U}\op{A}\op{V}} = \normM{\op{A}} 
\end{equation}
for all matrices $\op{A}$ and all unitary matrices~$\op{U}$ and~$\op{V}$. 

Now consider the perturbed Hamiltonian
\begin{equation}\label{eq:H}
	\Hperturbed = \Hs+\varepsilon\Hp
\end{equation}
with unperturbed Hamiltonian $\Hs$, perturbation $\Hp$, and perturbation strength~$\varepsilon \geq 0$. The eigenvalue equation of the perturbed Hamiltonian is
\begin{equation}\label{eq:EP}
	(\Hs+\varepsilon\Hp)\ket{\state_\ind} = \ev_\ind\ket{\state_\ind}
\end{equation}
with complex eigenvalues $\ev_\ind$ and right eigenstates $\ket{\state_\ind}$ normalized to unity: $\normV{\state_\ind} = 1$. Equation~(\ref{eq:EP}) can be written as
\begin{equation}\label{eq:EP2}
	\ket{\state_\ind} = \varepsilon\GF(\ev_\ind)\Hp\ket{\state_\ind}
\end{equation}
with the Green's function of the unperturbed Hamiltonian 
\begin{equation}\label{eq:GFdef}
	\GF(\ev) := (\ev\ident-\Hs)^{-1} \ ,
\end{equation}
where $\ident$ is the identity. Taking the vector norm on both sides of Eq.~(\ref{eq:EP2}) and using the normalization of the eigenstate gives
\begin{equation}\label{eq:EP3}
	1 = \varepsilon\normV{\GF(\ev_\ind)\Hp\state_\ind} \ .
\end{equation}
Exploiting the compatibility~(\ref{eq:compatible}) twice in Eq.~(\ref{eq:EP3}) yields
\begin{equation}\label{eq:EPineq}
	1 \leq \varepsilon\normM{\GF(\ev_\ind)}\,\normM{\Hp} \ .
\end{equation}
This inequality is valid both for the Frobenius and the spectral norms. However, the spectral norm is preferred as it gives the smallest bounds because of the general relation $\normspec{\op{A}} \leq \normFro{\op{A}}$. 

The derivation in Ref.~\cite{Wiersig22} is restricted to the case of an $\order\times\order$ Hamiltonian $\Hs$ to be at an EP of order~$\order$ with eigenvalue~$\evEP$. In this generalized eigenspace the matrix
\begin{equation}\label{eq:N}
	\op{N} := \Hs-\evEP\ident 
\end{equation}
is nilpotent of index $\order$; hence, $\op{N}^\order = 0$ but $\op{N}^{\order-1} \neq 0$. This property implies the expansion~\cite{Wiersig22}
\begin{equation}\label{eq:Heiss}
	\GF(\ev) = \frac{\ident}{\ev-\evEP} + \sum_{k=2}^\order \frac{\op{N}^{k-1}}{(\ev-\evEP)^k} \ .
\end{equation}
For $\ev\approx\evEP$ the contribution of the Green's function with $k=\order$ is the dominant one.
For generic perturbations with 
\begin{equation}\label{eq:generic}
\HL{\op{N}^{\order-1}\Hp\ket{\state_\ind} \neq 0}
\end{equation}
 we use only this dominant contribution in inequality~(\ref{eq:EPineq}) yielding
\begin{equation}
	|\ev_\ind-\evEP|^\order \leq \varepsilon \normspec{\Hp}\,\normspec{\op{N}^{\order-1}} \ .
\end{equation}
This gives reason to define the spectral response strength associated to the EP
\begin{equation}\label{eq:rca}
	\rca := \normspec{\op{N}^{\order-1}} \ .
\end{equation}
Sometimes it is necessary to distinguish EPs of different order~$\order$. Then we write $\rca_\order$ instead of $\rca$. With the definition in Eq.~(\ref{eq:rca}) one finally gets
\begin{equation}\label{eq:precentral}
	|\ev_\ind-\evEP|^\order \leq \varepsilon \normspec{\Hp}\, \rca  \ .
\end{equation}
This inequality is the central result of Ref.~\cite{Wiersig22}. It gives an upper bound for the energy eigenvalue change, or loosely speaking the energy splitting, $|\ev_\ind-\evEP|$, near an EP of order~$\order$. 
Importantly, the spectral response strength~$\rca$ is only a function of the unperturbed Hamiltonian $\Hs$. It is independent of the chosen orthonormal basis because the unitary invariance~(\ref{eq:unitary}) ensures the  invariance under a unitary transformation of~$\Hs$. However, a similarity transformation of~$\Hs$ in general does change~$\rca$.

The calculation of the spectral response strength in Eq.~(\ref{eq:rca}) is simple, in particular since the matrix $\op{N}^{\order-1}$ has rank 1 which follows from the nilpotency of~$\op{N}$ of index~$\order$. For rank-1 matrices the spectral norm~(\ref{eq:defspn}) and Frobenius norm~(\ref{eq:Fronorm}) give the same numerical values. Hence, the spectral response strength can be calculated more easily by
\begin{equation}\label{eq:rcafro}
	\rca = \normFro{\op{N}^{\order-1}} \ .
\end{equation}

The spectral response strength $\rca$ not only provides an upper bound for the energy eigenvalue change upon perturbation, it also gives upper bounds for the intensity response to excitation and the dynamic response to initial deviations from the EP eigenstate~\cite{Wiersig22}. 
Moreover, there is a relationship between $\rca$ and the Petermann factors of isolated eigenvalues resulting from a small generic perturbation of an EP$_\order$~\cite{Wiersig23}
\begin{equation}\label{eq:PFresult}
	\sqrt{\PF_\ind} = \frac{\rca}{\order|E_\ind-\evEP|^{\order-1}} \ , 
\end{equation}
where the Petermann factors are defined by 
\begin{equation}\label{eq:PF}
	\PF_\ind := \frac{\braket{R_\ind}{R_\ind}\braket{L_\ind}{L_\ind}}{|\braket{L_\ind}{R_\ind}|^2} \ ,
\end{equation}
with right eigenstates $\ket{R_\ind}$ and the corresponding left eigenstates~$\ket{L_\ind}$ of the Hamiltonian~$\op{H}$. Note that the Petermann factors of the involved eigenstates do not depend on the quantum number $\ind$ in this regime of small generic perturbations. 
From Eq.~(\ref{eq:PFresult}) it follows that the Petermann factor diverges when approaching the EP consistent with Ref.~\cite{Berry03}.
 
\section{Generalization to higher-dimensional Hilbert spaces}
\label{sec:general}
In this section we generalize the spectral response strength~$\rca$ to the case of an $m\times m$ Hamiltonian exhibiting an EP of order $\order \leq m$. Such a scenario \HL{appears, for instance, in the context of EPs in lattice systems~\cite{KKX22,KKM22}. The generic scenario} has been numerically treated in Ref.~\cite{Wiersig23} based on the natural assumption that the relation between $\rca$ and the Petermann factors near the EP in Eq.~(\ref{eq:PFresult}) is valid also in the general case. \HL{The validity of this assumption has been proven very recently in Ref.~\cite{HS23}}.
A general theory can be found, in principle, in the mathematical literature under the name  H{\"o}lder condition number~\cite{MBO97,Karow06}. However, the mathematical approach is difficult to access for physicists and not easy to apply to practical problems. 
Our approach is more elementary and presented in the physically relevant Green's function. We start with the general expansion of the $m\times m$ Green's function of the unperturbed Hamiltonian~$\Hs$~\cite{Kato66}
\begin{equation}\label{eq:GKato}
	\GF(\ev) = \sum_\ind\left[\frac{\projector_\ind}{\ev-\evn_\ind} + \sum_{k=2}^{\order_\ind} \frac{\nilpotent_\ind^{k-1}}{(\ev-\evn_\ind)^k}\right] \ .
\end{equation}
The sum over~$\ind$ covers the relevant part of the point spectrum including isolated energy eigenvalues and EPs. The sum over the integer~$k$ gives additional terms for EPs (order $\order_\ind \geq 2$) only. The operators~$\projector_\ind$ are projectors onto the generalized eigenspaces of the corresponding eigenvalues~$\evn_\ind$ with
\begin{equation}\label{eq:pp}
\projector_j\projector_\ind = \delta_{j\ind}\projector_\ind \ .
\end{equation}
In general, the $\projector_\ind$ are not orthogonal projectors, i.e., $\projector_\ind \neq \projector^\dagger_\ind$, reflecting the nonnormality of $\Hs$.   
The operators~$\nilpotent_\ind$ for a given EP of order $\order_\ind$ are nilpotent operators of index~$\order_\ind$ and defined by
\begin{equation}\label{eq:nilpotentN}
\nilpotent_\ind := \left(\Hs-\evn_\ind\ident\right)\projector_\ind \ .
\end{equation}
It holds $\projector_\ind \nilpotent_\ind = \nilpotent_\ind\projector_\ind = \nilpotent_\ind$.

Equation~(\ref{eq:GKato}) is general in contrast to the expansion in Eq.~(\ref{eq:Heiss}). An alternative approach based on left and right Jordan vectors~\cite{HKC22,SOE23} is not suitable for our purpose as it, in general, requires a similarity transformation of $\Hs$. Such a similarity transformation would be harmful as it may modify the spectral response strength that we want to calculate. 

It is possible to explicitly determine for a given Hamiltonian $\Hs$ the expansion of its Green's function in Eq.~(\ref{eq:GKato}) including all the operators~$\projector_\ind$ and $\nilpotent_\ind$ by using partial fraction decomposition. However, this is in general time-consuming and in fact not necessary. For our purpose it is sufficient to know the operator 
\begin{equation}\label{eq:W}
 \roneM := \nilpotent_\ind^{\order_\ind-1}  
\end{equation}
for the EP of interest. There is no need to know~$\nilpotent_\ind$ itself nor the projector~$\projector_\ind$. Our strategy is the following: We start with calculating analytically the Green's function of the unperturbed Hamiltonian in Eq.~(\ref{eq:GFdef}) by direct matrix inversion \HL{(for the cases where this is not possible we refer to the numerical scheme in Sec.~\ref{sec:residue})}. From the resulting Green's function we take, for an energy level $\evn_\ind = \evEP$, the leading-order contribution
\begin{equation}\label{eq:Gpole}
\GF(\ev) = \frac{\roneM}{\left(\ev-\evEP\right)^{\order_\ind}} \ ,
\end{equation}
where for the energy $\ev$ the following condition is assumed
\begin{equation}\label{eq:assumption}
|\ev-\evEP| \ll |\ev-\evn_j| \; \text{for}\, j \neq \ind \ ,
\end{equation}
i.e., the EP is isolated.
It this case we can safely replace $\ev-\evn_j$ for $j \neq \ind$ by $\evn_\ind-\evn_j$ which is then incorporated into $\roneM$. It is emphasized that the leading-order contribution within the more laborious partial fraction decomposition of the Green's function leads to the same result as in Eq.~(\ref{eq:Gpole}). This can be seen easily for the example
\begin{eqnarray}
\nonumber
\frac{1}{(E-\evn_\ind)(E-\evn_j)} & = & \frac{1}{(\evn_\ind-\evn_j)(E-\evn_\ind)} +\\ && \frac{1}{(\evn_j-\evn_\ind)(E-\evn_j)} .
\end{eqnarray}

Comparing Eq.~(\ref{eq:Gpole}) with expansion~(\ref{eq:Heiss}) uncovers that the operator~$\roneM$ plays the role of the operator~$\op{N}^{\order-1}$. All arguments leading to Eqs.~(\ref{eq:rca}) and~(\ref{eq:precentral}) carry over. Assuming again a generic perturbation [Eq.~(\ref{eq:generic})] results in
\begin{equation}\label{eq:rcageneral}
	\rca := \normspec{\roneM} \ .
\end{equation}
Again, since $\nilpotent_\ind$ is nilpotent of index $\order_\ind$, the operator $\roneM$ is of rank 1. Hence, the spectral response strength can be calculated more conveniently by the Frobenius norm
\begin{equation}\label{eq:rcageneralFro}
	\rca = \normFro{\roneM} \ .
\end{equation}

When considered superficially, the final formulas for the generalized theory, Eqs.~(\ref{eq:Gpole}), (\ref{eq:rcageneral}), and (\ref{eq:rcageneralFro}), appear to be straightforward generalizations of the formulas of the special theory, Eqs.~(\ref{eq:rca}) and (\ref{eq:rcafro}) in the previous section. However, the way the calculation is performed is entirely different. 
In the special theory, first the nilpotent operator~$\op{N}$ is computed via Eq.~(\ref{eq:N}). Then $\op{N}^{\order-1}$ is calculated by matrix powers and plugged into Eqs.~(\ref{eq:rca}) or~(\ref{eq:rcafro}) to obtain the spectral response strength.
In the general theory presented here, the full Green's function in Eq.~(\ref{eq:GFdef}) is first determined by matrix inversion. Then the leading-order contribution of the Green's function for the EP of interest is identified as in Eq.~(\ref{eq:Gpole}). From this contribution, the operator~$\roneM$ is extracted and inserted into Eqs.~(\ref{eq:rcageneral}) or~(\ref{eq:rcageneralFro}) to  obtain the spectral response strength.

\section{Divergence near higher-order EPs}
\label{sec:divergence}
What happens if we push our theory to a regime where the assumptions~(\ref{eq:assumption}) fail? Consider the interesting case where one EP of order~$\order$ moves under variation of parameters along an exceptional surface approaching an EP of higher order $\order+k$ with integer $k\geq 1$. We denote the corresponding operators by $\roneM_\order$ and $\roneM_{\order+k}$ and the eigenvalues by $\evEPk{\order}$ and $\evEPk{\order+k}$. The replacement $\ev-\evn_j$ for $j \neq \ind$ by $\evn_\ind-\evn_j$ discussed in the context of Eq.~(\ref{eq:Gpole}) leads to the proportionality
\begin{equation}\label{eq:Wtransition}
	\roneM_\order \propto \frac{\roneM_{\order+k}}{\left(\evEPk{\order}-\evEPk{\order+k}\right)^{k}} \ .
\end{equation}
From this equation and Eq.~(\ref{eq:rcageneral}) it follows
\begin{equation}\label{eq:rcascaling}
	\rca_\order\left|\evEPk{\order}-\evEPk{\order+k}\right|^k \propto \rca_{\order+k} \ .
\end{equation}
From this relation we learn that the spectral response strength of the lower-order EP diverges to infinity as a higher-order EP with nonzero spectral response strength is approached. Clearly, this can only happen if $m > \order$, i.e., if the Hilbert space dimension exceeds the order of the lower-order EP.
For fixed perturbation strength~$\varepsilon > 0$ the divergence is artificial as it is caused by the invalidity of the assumptions~(\ref{eq:assumption}) for too small detuning $|\evEPk{\order}-\evEPk{\order+k}|$. In such an extreme situation with several competing poles in the expansion in Eq.~(\ref{eq:GKato}) it is not possible to describe the spectral response of the system by a single quantity. 

Nevertheless, a continuously increasing spectral response strength~$\rca$ is not artificial but natural and even essential for describing the physics near a higher-order EP correctly. To see this, imagine $\rca_{\order+k} \neq 0$, a perturbation strength~$\varepsilon$ and a detuning such that the assumptions~(\ref{eq:assumption}) are not valid. According to Eq.~(\ref{eq:rcascaling}) the spectral response strength~$\rca_\order$ is large. Now, for fixed detuning let us reduce $\varepsilon$. When $\varepsilon$ is sufficiently small the assumptions~(\ref{eq:assumption}) are valid again as the perturbed eigenvalues~$\ev_\ind$ stay close enough to $\evEP$. As the spectral response strength is independent of $\varepsilon$, a previous large value is needed to describe the enhanced spectral response near the higher-order EP.

Assuming that the matrix elements of $\Hs$ stay finite, we can see from the definition of $\roneM$ in Eqs.~(\ref{eq:nilpotentN}) and~(\ref{eq:W}) that the origin of the divergence of $\rca$ is a diverging projector~$\projector_\ind$. This is the essential difference to Ref.~\cite{Wiersig22} where the projector is just the identity~$\ident$.

For systems without gain, i.e., passive systems, Ref.~\cite{Wiersig22b} has derived for the special case $m = \order$ an upper bound for the spectral response strength 
\begin{equation}\label{eq:rcapassve}
	\rca \leq \left(\sqrt{2\order}|\imagc{\evEP}|\right)^{\order-1} \ .
\end{equation}
This is a considerable limitation for sensor applications as a small linewidth $\lw = 2|\imagc{\evEP}|$ implies a small splitting for a given perturbation. Obviously, inequality~(\ref{eq:rcapassve}) cannot be true for $m > \order$ with a diverging $\rca$. Going through the derivation in Ref.~\cite{Wiersig22b} reveals that the origin of the failure of inequality~(\ref{eq:rcapassve}) is the projector~$\projector_\ind \neq \ident$.

Finally, it is mentioned that another kind of transition between EPs of different order is possible. Instead of the coalescence of eigenenergies discussed above it can also happen that the leading-order contribution corresponding to, let us say, an EP of order $\order+k$ vanishes, $\roneM_{\order+k} = 0$, under a parameter variation. Then the leading-order contribution of another EP, let us say of order $\order$, takes over. If this happens without any change of eigenenergies then Eq.~(\ref{eq:rcascaling}) still applies since $\rca_{\order+k} = 0$.

\section{A toy model}
\label{sec:toymodel}
For illustration purposes, we introduce a minimal model, simple enough to get analytical results even by hand, but showing the relevant effects clearly. We consider the $3\times 3$ Hamiltonian
\begin{equation}\label{eq:H33}
\Hs = \left(\begin{array}{ccc}
\Eb & B & 0\\
0   & \Ea & A\\
0   & 0 & \Ea\\
\end{array}\right) \ ,
\end{equation}
with complex parameters $\Ea$, $\Eb$, $A \neq 0$, and $B \neq 0$. The eigenvalues of $\Hs$ are $\Ea$ (with algebraic multiplicity of two) and $\Eb$. 

With the definition in Eq.~(\ref{eq:GFdef}) the Green's function is computed by matrix inversion 
\begin{equation}\label{eq:GF33}
\GF(E) = \left(\begin{array}{ccc}
	\frac{1}{E-\Eb} & \frac{B}{(E-\Eb)(E-\Ea)} & \frac{AB}{(E-\Eb)(E-\Ea)^2}\\
	0   & \frac{1}{E-\Ea} & \frac{A}{(E-\Ea)^2}\\
	0   & 0 & \frac{1}{E-\Ea}\\
\end{array}\right) \ .
\end{equation}	
The reader may crosscheck that this Green's function is of the form of the expansion in Eq.~(\ref{eq:GKato}) by exploiting partial fraction decomposition. However, this is, as stressed in the previous section, not a precondition for the computation of the spectral response strength.

Let us first focus on the generic case $\Eb \neq \Ea$. Here, the leading-order contribution at $\ev = \Ea$ according to Eq.~(\ref{eq:Gpole}) is of order $\order = 2$ and with the replacement of $\ev-\Eb$ by $\Ea-\Eb$ 
\begin{equation}\label{eq:W2}
\roneM = \left(\begin{array}{ccc}
0 & 0 & \frac{AB}{\Ea-\Eb}\\
0 & 0 & A\\
0 & 0 & 0\\
\end{array}\right) \ .
\end{equation}	
Note that the same result is obtained by partial fraction decomposition of the Green's function in Eq.~(\ref{eq:GF33}). With Eq.~(\ref{eq:rcageneralFro}) the spectral response strength of this second-order EP with eigenvalue $\evEPk{2} = \Ea \neq \Eb$ is
\begin{equation}\label{eq:rca2}
\rca_2 = |A|\sqrt{1+\frac{|B|^2}{|\Eb-\Ea|^2}} \ .
\end{equation}	
In the special case $\Eb = \Ea$ the leading-order contribution in Eq.~(\ref{eq:GF33}) is, according to Eq.~(\ref{eq:Gpole}), of order $\order = 3$ and
\begin{equation}\label{eq:W3}
	\roneM = \left(\begin{array}{ccc}
		0 & 0 & AB\\
		0 & 0 & 0\\
		0 & 0 & 0\\
	\end{array}\right) \ .
\end{equation}	
With Eq.~(\ref{eq:rcageneralFro}) the spectral response strength of this third-order EP with eigenvalue $\evEPk{3} = \Eb$ is
\begin{equation}\label{eq:rca3}
	\rca_3 = |A||B| \ .
\end{equation}	

To verify our results numerically we consider the Hamiltonian in Eq.~(\ref{eq:H}) with a perturbation 
\begin{equation}\label{eq:H33p}
	\Hp = \left(\begin{array}{ccc}
		0 & 0 & 0\\
		0 & 0 & 0\\
		1/\sqrt{2} & 1/\sqrt{2} & 0\\
	\end{array}\right) \ .
\end{equation}
This Hamiltonian represents a generic perturbation that leads to $\order$th root energy splitting. It is normalized to unity, $\normspec{\Hp} = 1$, and, hence, the strength of the perturbation is here fully determined by~$\varepsilon$. \HL{Note that any randomly chosen perturbation $\Hp$ would be equally good for our purpose as long as it is generic, see Eq.~(\ref{eq:generic}).}
A variation of the detuning~$|\Eb-\Ea|$ can be regarded as a nongeneric perturbation that leaves the system on an EP of the same order. It therefore shifts the system on a curve in an exceptional surface.

\begin{figure}[ht]
	\includegraphics[width=0.95\columnwidth]{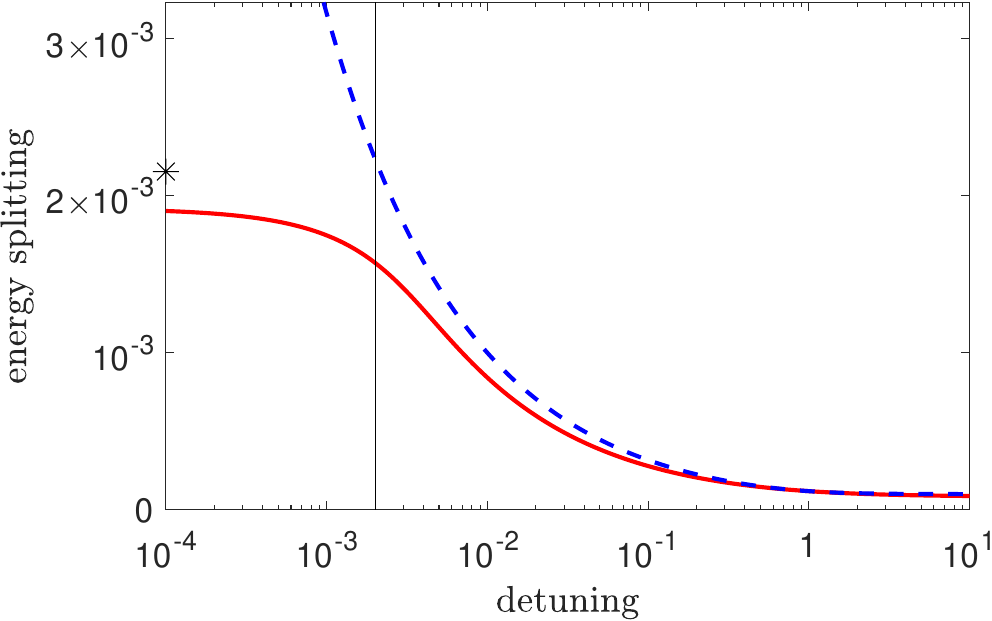}
	\caption{Energy splitting and corresponding bounds (both dimensionless) for the Hamiltonian in Eq.~(\ref{eq:H33}) and perturbation in Eq.~(\ref{eq:H33p}) vs detuning $|\Eb-\Ea|$ in a semilogarithmic scale. The thin vertical line marks the detuning $2\times 10^{-3}$ focused on in Fig.~\ref{fig:splittingeps}. The solid curve is the change of the eigenvalue nearest to $\Ea$ computed numerically from the eigenvalue equation. The dashed curve is the upper bound in Eq.~(\ref{eq:precentral}) with the spectral response strength in Eq.~(\ref{eq:rca2}) for the second-order EP. The star marks the upper bound in Eq.~(\ref{eq:precentral}) with the spectral response strength in Eq.~(\ref{eq:rca3}) for the third-order EP at zero detuning (horizontally shifted to the smallest finite detuning visible in the figure). The parameters are $\varepsilon = 10^{-8}$, $A = -1 = B$, and $\imagc{\Ea} = \imagc{\Eb} = 0$.}
	\label{fig:splitting}
\end{figure}

Figure~\ref{fig:splitting} shows the energy splitting and the upper bounds in Eq.~(\ref{eq:precentral}) when the system is moving along this exceptional surface. The upper bound with the spectral response strength in Eq.~(\ref{eq:rca2}) appears to be a tight one for detunings above $2\times 10^{-3}$, i.e., for 99.98 percent of the considered part of the exceptional surface. Only for detunings below $2\times 10^{-3}$ the bound does not seem to be tight. In this regime the energy splitting is of similar size as the detuning or even larger, and hence the assumptions~(\ref{eq:assumption}) are no longer justified. Approaching zero detuning the spectral response strength in Eq.~(\ref{eq:rca2}) diverges. Note that the upper bound in Eq.~(\ref{eq:precentral}) is still fulfilled. 
The divergence of the upper bound is, at first glance, reminiscent to the behavior of energy eigenvalues near so-called divergent EPs~\cite{SHZ19,YZZ23}. However, the eigenvalue changes here remain finite. This must be the case because for zero detuning the upper bound in Eq.~(\ref{eq:precentral}) has to hold for $\order = 3$ with the spectral response strength in Eq.~(\ref{eq:rca3}).
The spectral response strength of the second-order EP in Eq.~(\ref{eq:rca2}) scales close to the third-order EP ($\Eb \approx\Ea$) as
\begin{equation}
	\rca_2 \approx \frac{|A||B|}{|\Eb-\Ea|} = \frac{\rca_3}{\left|\evEPk{2}-\evEPk{3}\right|}\ ,
\end{equation}	
where the spectral response strength in Eq.~(\ref{eq:rca3}) has been used. This is in full agreement with the prediction in Eq.~(\ref{eq:rcascaling}) for $\order = 2$ and $k = 1$.

\begin{figure}[ht]
	\includegraphics[width=0.95\columnwidth]{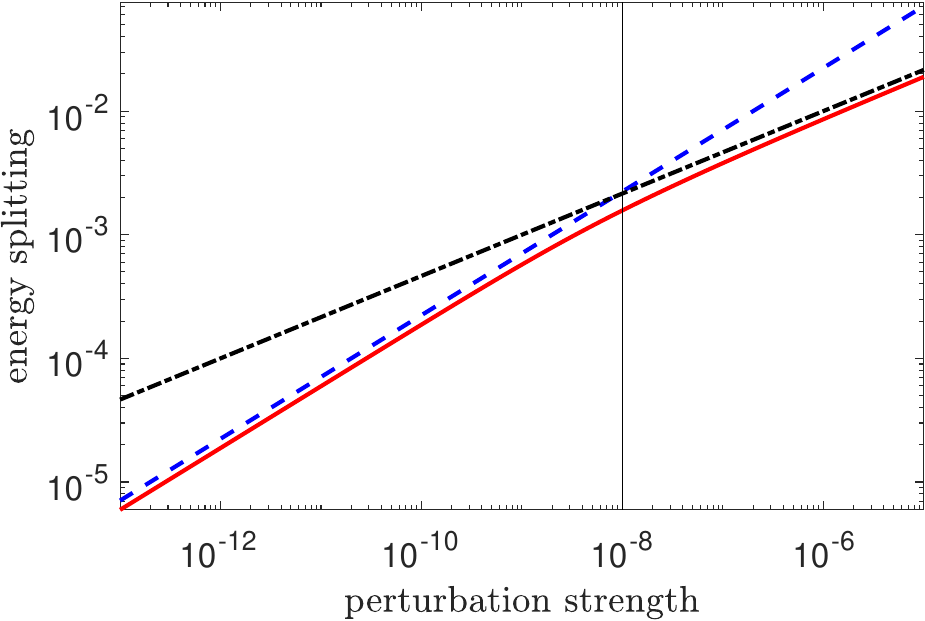}
	\caption{Energy splitting and corresponding bounds (both dimensionless) for the Hamiltonian in Eq.~(\ref{eq:H33}) and perturbation in Eq.~(\ref{eq:H33p}) vs perturbation strength in a double logarithmic scale. The thin vertical line marks the perturbation strength $\varepsilon = 10^{-8}$ used in Fig.~\ref{fig:splitting}. The solid curve is the change of the eigenvalue nearest to $\Ea$ computed numerically from the eigenvalue equation. The dashed curve is the upper bound in Eq.~(\ref{eq:precentral}) with the spectral response strength in Eq.~(\ref{eq:rca2}) for the second-order EP. The dash-dotted curve marks the upper bound in Eq.~(\ref{eq:precentral}) with the spectral response strength in Eq.~(\ref{eq:rca3}) for the third-order EP at zero detuning. The parameters are  $A = -1 = B$, $\Eb-\Ea = 2\times 10^{-3}$, and $\imagc{\Ea} = \imagc{\Eb} = 0$.}
	\label{fig:splittingeps}
\end{figure}

From Fig.~\ref{fig:splitting} we have learned that the spectral response strength does not give a tight bound for detunings below $2\times 10^{-3}$. Clearly, this statement depends on the given parameters, in particular on the perturbation strength~$\varepsilon$. Figure~\ref{fig:splittingeps} shows the splitting and the bounds for a detuning of $2\times 10^{-3}$ as function of~$\varepsilon$. Obviously, the bounds in Eq.~(\ref{eq:precentral}) appear here as straight lines with slope $1/2$ and $1/3$, respectively. The spectral response strengths themselves are independent of $\varepsilon$. For~$\varepsilon$ above $10^{-8}$ the spectral response strength of the second-order EP seems not to lead to a tight bound. It is larger than that of the spectral response strength of the third-order EP nearby (such a \quoting{wrong} scaling has been observed previously in a photonic waveguide-microring system~\cite{KGK23}). However, for smaller perturbation strength it is the other way round. It is clear from inequality~(\ref{eq:precentral}) that a large $\rca$ is needed to get an accurate bound for these small values of $\varepsilon$. It is important to mention that in Fig.~\ref{fig:splittingeps} both spectral response strengths together offer a comprehensive picture of the splitting. To conclude, the divergence of the spectral response strength for zero detuning does not hinder the proper description of the energy splitting even for small detuning as long as the perturbation strength is sufficiently small.
	
\HL{The dominant contribution to the Green's function in Eq.~(\ref{eq:GF33}) for $\ev$ close to but different from $\Ea$ and $\Eb$ is its top right matrix element which can be written as
\begin{eqnarray}
\nonumber
\frac{AB}{(E-\Eb)(E-\Ea)^2} & = &\frac{AB}{(\Ea-\Eb)(E-\Ea)^2}\\
\label{eq:domcontri}
 & - &\frac{AB}{(\Ea-\Eb)(E-\Eb)(E-\Ea)} \ .
\end{eqnarray}
The first term on the right-hand side is the one that enters the calculation of the spectral response strength of the second-order EP for $\Ea \neq \Eb$. This term diverges to infinity for $\Eb\to\Ea$. However, the spectral response to perturbations (in terms of energy eigenvalue changes) remains finite as the second term on the right-hand side of Eq.~(\ref{eq:domcontri}), which has a pole at the energy of the isolated eigenvalue, cancels the divergence. This is analog to the cancellation of divergences in the case of two isolated eigenvalues coalescing at a second-order EP~\cite{YSS11}.
}

It is also illuminating to interpret the results in Fig.~\ref{fig:splittingeps} in terms of pseudospectra.  These generalized spectra are an alternative way to study the sensitivity of non-normal matrices subjected to perturbations~\cite{TE05}. Applications in optics and photonics can be found in Refs.~\cite{MGT14,ZE19,Makris21,KKX22,KKM22}.
Given a positive number $\pseps$, the $\pseps$-pseudospectrum of a non-normal matrix~$\Hs$ can be defined as the subset of the complex plane
\begin{equation}\label{eq:ps1}
	\pseudospectrum_{\pseps} := \{\ev\in\C: \normM{\GF(\ev)} > 1/\pseps\}
\end{equation}
with Green's function~(\ref{eq:GFdef}) and arbitrary matrix norm $\normM{\cdot}$, which we fix here to be the spectral norm. The pseudospectrum of $\Hs$ includes not only the spectrum of $\Hs$ as poles of $\GF(\ev)$ but also the eigenvalues resulting from a normalized perturbation, $\normspec{\Hp} = 1$, with perturbation strength $\pseps$; as can be understood from inequality~(\ref{eq:EPineq}).
The spectral response strength~$\rca$ determines the radius~$\sqrt[\order]{\pseps\rca}$ of the pseudospectrum-disk (the $\pseps$-pseudospectral radius~\cite{TE05}) close to the EP$_\order$ for sufficiently small~$\pseps$~\cite{Wiersig22}.

\begin{figure}[ht]
	\includegraphics[width=0.8\columnwidth]{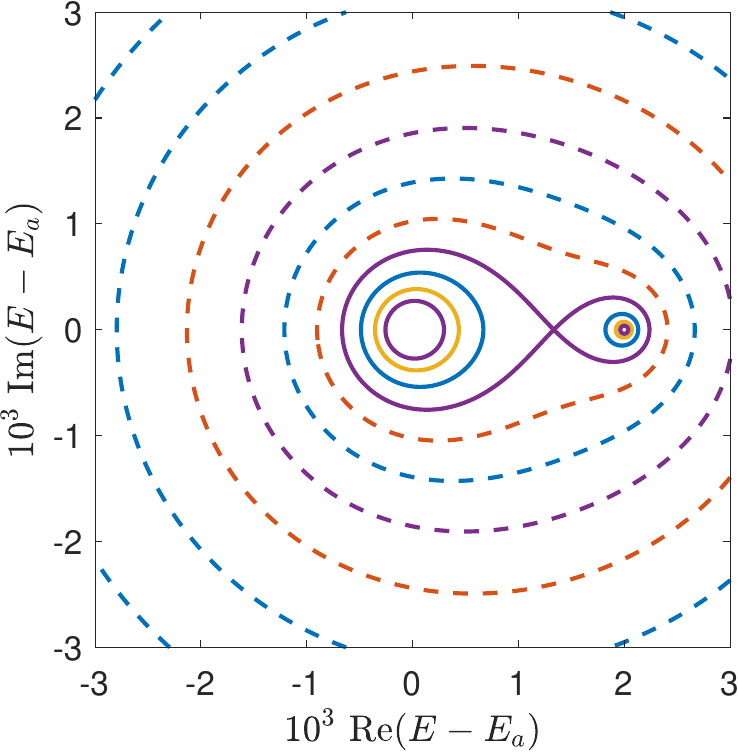}
	\caption{Contour plot of the pseudospectrum [Eq.~(\ref{eq:ps1})] of the Hamiltonian in Eq.~(\ref{eq:H33}) in the complex energy plane (dimensionless) with parameters as in Fig.~\ref{fig:splittingeps}. The dashed curves are the isolines of the spectral norm of the Green's function for $\pseps = 10^{\crit+0.3}$, $10^{\crit+0.6}$ etc. with $\crit = -8.9262$. The solid curves are the isolines for $\pseps = 10^{\crit}$, $10^{\crit-0.3}$ , $10^{\crit-0.6}$ etc.}
	\label{fig:pseudospectrum}
\end{figure}

Figure~\ref{fig:pseudospectrum} shows the pseudospectrum of the toy model with the same parameters as before. The pole related to the EP$_2$ with eigenvalue $\Ea$ is located in the center, and the pole related to the isolated eigenvalue~$\Eb$ is to the right of it. For small~$\pseps$ the isolines of the Green's function are small circles around the respective pole. Hence, the response of the EP$_2$ and the isolated eigenvalue to perturbations can be regarded as being independent. For the same $\pseps$, the circle is larger for the EP$_2$ signaling a larger response if compared to the individual eigenvalue. This situation corresponds to the low $\varepsilon$-regime in Fig.~\ref{fig:splittingeps}.
At the critical $\pseps = 10^\crit$ with $\crit = -8.9262$ the isolines related to the EP$_2$ and the isolated eigenvalue touch each other forming a separatrix-like curve. This transition fits into the transition region in Fig.~\ref{fig:splittingeps}. For a larger $\pseps$ a single closed isoline prevails which becomes more and more circular shaped as~$\pseps$ is increased further. The joint response of the EP$_2$ and the isolated eigenvalue become indiscernible from that of a single EP$_3$. In Fig.~\ref{fig:splittingeps} this corresponds to the regime of large perturbations where the energy splitting is well described by the spectral response strength of the EP$_3$. 
To conclude, the pseudospectrum in Fig.~\ref{fig:pseudospectrum} gives an intuitive understanding of why the energy splitting as function of perturbation strength in Fig.~\ref{fig:splittingeps} shows a transition from an EP$_2$ to an EP$_3$.

Finally, let us have a brief look onto the special case $\Eb = \Ea$ with variable parameter $B$.  For $B\neq 0$, even if $B$ is very small, the Hamiltonian in Eq.~(\ref{eq:H33}) is at an EP$_3$ with spectral response strength $\rca_3 = |A||B|$. Hence, varying the parameter $B$ moves the system along an exceptional surface of third order. For $B=0$, however, the Hamiltonian possesses an EP$_2$ degenerated with the former isolated eigenvalue. The spectral response strength of the second-order EP is $\rca_2 = |A|$. Hence, there is an abrupt transition without a diverging spectral response strength. However, Eq.~(\ref{eq:rcascaling}) is still correct as $\rca_3 = 0$ in the limit $B\to 0$.

\section{A realistic example}
\label{sec:example}
After we have discussed the toy model in Sec.~\ref{sec:toymodel} we now turn our attention to a more realistic example studied in the context of transport of chirality in a deformed microdisk cavity~\cite{LWS18}. Deformed microcavities have been studied a lot in the field of non-Hermitian photonics and wave chaos~\cite{CW15}. In the latter case, one is interested in the correspondence of wave properties and ray dynamics in phase space. The relevant wave properties in the deformed cavity in Ref.~\cite{LWS18} are successfully modeled by the $4\times 4$-Hamiltonian 
\begin{equation}\label{eq:HTC}
\Hs = \left(\begin{array}{cccc}
\freqisland & V            & 0            & 0           \\
V 	        & \freqchaotic & A            & 0           \\
0           & B            & \freqchaotic & V           \\
0			& 0			   & V            & \freqisland \\ 
	\end{array}\right) \ .
\end{equation}
This Hamiltonian describes the dynamics of a four-state system. In the optics context, the energy eigenstates are denoted optical modes and the eigenenergies are the eigenfrequencies of the optical modes. If we ignore the coupling elements $V$, $A$, and $B$ for a moment then we can say that two modes with complex frequency~$\freqisland$ are localized in a so-called regular-island chain, a nonchaotic part of the ray-dynamical phase space. One regular-island chain corresponds to clockwise propagation in real space and the other one to counterclockwise propagation. The other two counterpropagating modes with complex frequency~$\freqchaotic$ are localized in the chaotic region of phase space between the two regular-island chains. The coupling element~$V$ describes the dynamical tunneling between a regular-island chain and the surrounding chaotic region. The backscattering between clockwise and counterclockwise propagating waves in the chaotic region is asymmetric, $|A| \neq |B|$, see Ref.~\cite{Wiersig18b} for a review of asymmetric backscattering. The asymmetry originates physically from a waveguide that is connected to the cavity in a spatially asymmetric manner. 

The eigenvalues of the Hamiltonian in Eq.~(\ref{eq:HTC}) are the frequencies of the optical modes. A short calculation gives 
\begin{eqnarray}
\nonumber
\Omega_{\pm,\sigma} & = & \frac{\freqisland+\freqchaotic+\sigma\sqrt{AB}}{2}\\
\label{eq:evTC}
& & \pm \sqrt{V^2 + \left(\frac{\freqisland-\freqchaotic-\sigma\sqrt{AB}}{2}\right)^2}	
\end{eqnarray}
where the sign in front of the square root and the quantity $\sigma = \pm 1$ label up to four eigenfrequencies. 
Here, we are interested in the special case $A\neq 0$, $B = 0$ (the case $A=0$, $B\neq 0$ is similar and not discussed here) which gives two distinct eigenvalues
\begin{equation}\label{eq:evTCEP}
	\Omega_\pm = \frac{\freqisland+\freqchaotic}{2}\pm \sqrt{V^2 + \left(\frac{\freqisland-\freqchaotic}{2}\right)^2}	\ .
\end{equation}
If the square root is nonzero, then the system possess two second-order EPs and otherwise a fourth-order EP. The argument in the square root can be considered as a complex variable which can be varied to move the system within an exceptional surface of second order towards an EP of fourth order.
Following the procedure in Sec.~\ref{sec:general} the spectral response strength of the two second-order EPs turn out to be
\begin{equation}\label{eq:rcaTCEP2}
\rca_{2,\pm} = |A| \frac{|V|^2 + |\Omega_\pm-\freqisland|^2}{|\Omega_\mp-\Omega_\pm|^2} \ .
\end{equation}
The spectral response strength for the fourth-order EP is
\begin{equation}\label{eq:rcaTCEP4}
	\rca_4 = |A| (|V|^2 + |\Omega_\pm-\freqisland|^2) 
\end{equation}
where the two eigenvalues $\Omega_\pm$ are coalesced to $(\freqisland+\freqchaotic)/2$. It can be clearly seen that when the two second-order EPs merge into a fourth-order EP, the response strength in Eq.~(\ref{eq:rcaTCEP2}) diverges in accordance with Eq.~(\ref{eq:rcaTCEP4}) and Eq.~(\ref{eq:rcascaling}) for $k = 2$.

\section{Relation to the divergence of the Petermann factor}
\label{sec:Petermann}
The divergence of the spectral response strength in Eq.~(\ref{eq:rcascaling}) is reminiscent to that of the Petermann factor~$\PF_\ind$ approaching an EP in Eq.~(\ref{eq:PFresult}). In fact, the behavior is exactly the same if we understood $\sqrt{\PF_\ind}$ as the response strength of isolated eigenvalue and choose $k = \order-1$. 
%
This analogy between Petermann factor and spectral response strength can be made more rigorous. To do so, we consider an eigenstate $\ket{\state_\ind}$ with eigenvalue~$\evn_\ind$ not an EP. The Green's function from Eq.~(\ref{eq:GKato}) simplifies to $\GF(\ev) = \projector_\ind/(
\ev-\evn_\ind)$. Plugging it into inequality~(\ref{eq:EPineq}) yields
\begin{equation}
	|\ev_\ind-\evn_\ind| \leq \varepsilon \normspec{\Hp}\,\normspec{\projector_\ind} \ .
\end{equation}
This is the Bauer-Fike theorem for the change of an isolated eigenvalue under perturbation~\cite{BF60}. Comparison with inequality~(\ref{eq:precentral}) shows that $\normspec{\projector_\ind}$ plays for the isolated eigenvalue the same role as the spectral response strength~$\rca$ for an EP.

The Bauer-Fike theorem is a standard result in perturbation theory of nonnormal matrices. It is also known that $\normspec{\projector_\ind}$ is related to the Petermann factor in Eq.~(\ref{eq:PF}), see, e.g., Ref.~\cite{BF60}, even though the name \quoting{Petermann factor} is not used in the mathematical literature. For completeness we present a short derivation of the relation between  $\normspec{\projector_\ind}$ and~$\PF_\ind$. To do so, we write the projector in terms of right and left eigenstates as
\begin{equation}\label{eq:pK}
	\projector_\ind = \frac{\ket{R_\ind}\bra{L_\ind}}{\braket{L_\ind}{R_\ind}} \ .
\end{equation}
This projector maps the right eigenstate~$\ket{\state_\ind}$ onto itself. The conditions in Eq.~(\ref{eq:pp}) for the projectors are fulfilled as $\braket{L_j}{R_\ind} = 0$ if $j\neq\ind$. 
The  projector in Eq.~(\ref{eq:pK}) clearly has rank 1 and therefore with Eq.~(\ref{eq:Fronorm})
\begin{equation}\label{eq:PPTr}
\normspec{\projector_\ind}^2 = \normFro{\projector_\ind}^2 = \trace{(\projector_\ind^\dagger \projector_\ind)} \ .
\end{equation}
With the normalization $\normV{R_\ind} = 1 = \normV{L_\ind}$ and with $\trace{(\ket{a}\bra{b})} = \braket{b}{a}$ one gets
\begin{equation}
\trace{(\projector_\ind^\dagger \projector_\ind)} = \frac{1}{|\braket{L_\ind}{R_\ind}|^2} \ .
\end{equation}
Using this equation together with Eq.~(\ref{eq:PPTr}) and the definition of the Petermann factor in Eq.~(\ref{eq:PF}) yields
\begin{equation}
\normspec{\projector_\ind} = \sqrt{\PF_\ind} \ .
\end{equation}
Hence, $\sqrt{\PF_\ind}$ can be considered as the spectral response strength of the isolated eigenvalue with quantum number~$\ind$. The divergence of this quantity approaching an EP is then just a particular case of the divergence of the spectral response strength of an EP approaching a higher-order EP in Eq.~(\ref{eq:rcascaling}). The divergence of the Petermann factor is also artificial in the sense that the change of the energy eigenvalues remains in general finite. But it is also essential in the sense that the strong response to small perturbations near a higher-order EP is correctly described. 

\HL{For the spectral response strength and the Petermann factor, the divergence indicates a structural change of the eigenvalues (from isolated eigenvalue to EP or from lower-order EP to higher-order EP) which can be directly observed in experiments. This conclusion is consistent with the transition from Lorentzian spectral lineshape to squared Lorentzian discussed in Ref.\cite{YSS11}.}

\section{Numerical scheme based on residue calculus}
\label{sec:residue}
When the dimension of the Hilbert space exceeds four an analytical calculation of the Green's function~in Eq.~(\ref{eq:GFdef}) is in general becoming too cumbersome. Therefore, we introduce here a numerical method based on Eqs.~(\ref{eq:GKato}), (\ref{eq:W}), and~(\ref{eq:rcageneralFro}) for this scenario. For a given EP of order~$\order_\ind$ and eigenvalue~$\evEP = \evn_\ind$ we apply residue calculus to the expansion in Eq.~(\ref{eq:GKato}), see also Ref.~\cite{Kato66}, to obtain
\begin{equation}\label{eq:residue}
	\roneM = \nilpotent_\ind^{\order_\ind-1} = \frac{1}{2\pi i}\oint_{C}d\ev\,(\ev-\evEP)^{\order_\ind-1}\GF(\ev) \ ,
\end{equation}
with $C$ being a simple closed integration path in the complex energy plane that separates the EP eigenvalue $\evEP$ under consideration from the other eigenvalues. For convenience a circle of radius~$r_C$ is chosen. The Green's function $\GF(E)$ is calculated numerically for each energy~$E$ by matrix inversion and is integrated along the path $C$ in Eq.~(\ref{eq:residue}). Finally, the spectral response strength $\rcanum$ is numerically determined by Eq.~(\ref{eq:rcageneralFro}).

\begin{figure}[ht]
	\includegraphics[width=0.95\columnwidth]{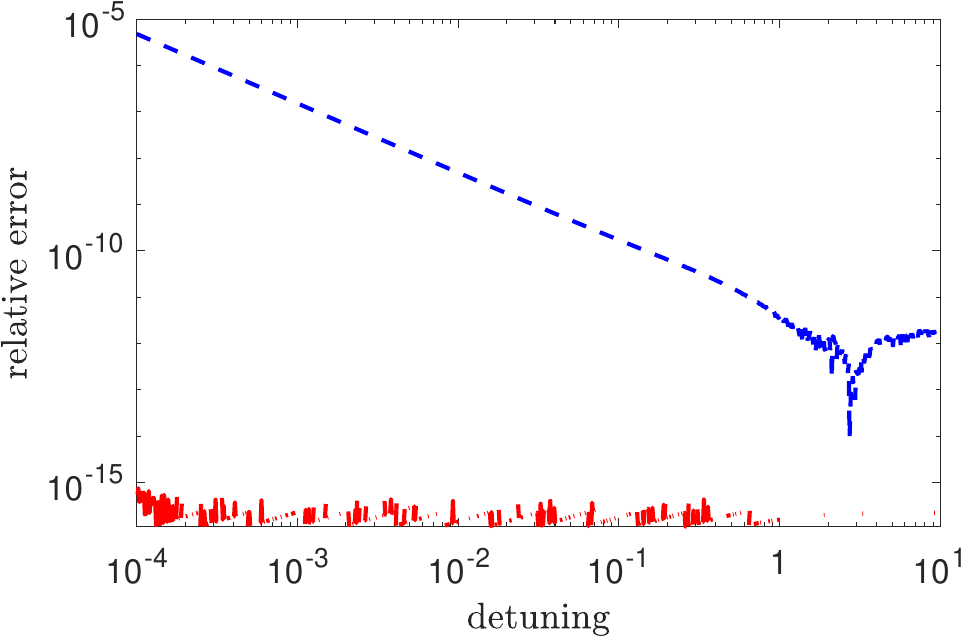}
	\caption{Relative error $|\rcanum-\rca|/\rca$ vs detuning $|\Eb-\Ea|$ (both are dimensionless) for the Hamiltonian in Eq.~(\ref{eq:H33}) in a double logarithmic scale; see Fig.~\ref{fig:splitting}. The lower curve, interrupted by numerically exact data points, marks the numerically determined spectral response strength $\rcanum$ using residue calculus in Eq.~(\ref{eq:residue}). The upper, dashed curve is based on the relation of Petermann factor and~$\rca$ in Eq.~(\ref{eq:PFresult}); see Ref.~\cite{Wiersig23}. The physical parameters are $A = -1 = B$, and $\imagc{\Ea} = \imagc{\Eb} = 0$. The parameter of the methods are  $r_C = 10^{-11}$ and $\eta = 10^{-21}$.}
	\label{fig:xinum}
\end{figure}

Figure~\ref{fig:xinum} shows for the toy model discussed in Sec.~\ref{sec:toymodel} the error $|\rcanum-\rca|$ of this method relative to the exact spectral response strength $\rca$ from Eq.~(\ref{eq:rca2}). With the relative error below $10^{-15}$ in the entire regime of considered detunings, the agreement between exact~$\rca$ and numerically determined $\rcanum$ is extremely good. The missing data points indicate that here the numerical result using double-precision floating-point arithmetic in MATLAB is even exact within machine precision. 

The dashed curve in Fig.~\ref{fig:xinum} shows the relative error of the numerical scheme based on the relation of the Petermann factor and $\rca$ in Eq.~(\ref{eq:PFresult}). According to the scheme  introduced in Ref.~\cite{Wiersig23} one keeps the Petermann factor finite by adding a tiny random perturbation of size $\eta$ to the Hamiltonian. This random perturbation has an influence on the relative error which can be reduced by adapting the random perturbation separately for each value of the system parameters. Here, we simply choose a uniform size of the random perturbation $\eta = 10^{-21}$ close to the optimum for the full interval of considered detunings.
We clearly see that this approach performs sufficiently well for most purposes but our novel approach based on residue calculus is far superior in terms of accuracy. It is, however, one to two orders of magnitude slower. 
Moreover, the superiority in terms of accuracy is reduced in a noisy environment (not shown), for instance for the random EPs studied in Ref.~\cite{Wiersig23}.

\section{Conclusion}
\label{sec:conclusion}
We have extended the theory of the spectral response strength to EPs of arbitrary order and to arbitrary Hilbert-space dimension. 
The theory gives an accurate bound for the changes of the energy eigenvalues for large parameter regions (in a given exceptional surface) in very good agreement with numerical solutions of the eigenvalue problem. Only very close to higher-order EPs the spectral response strength shows a divergence to infinity that is in general not accompanied by a divergence of energy eigenvalue changes. However, the divergence signals a phase transition and is essential for describing the strong response near higher-order EPs. This intriguing phenomenon has been demonstrated for two example systems and has been linked to the divergences (and their cancellation) of the Petermann factors of two energy eigenstates coalescing at a second-order EP.  

We have combined the general theory with residue calculus to establish an accurate and robust numerical scheme for the computation of the spectral response strength for the general case. For an exactly-solvable model, we have demonstrated an excellent agreement between theoretical and numerical results. 

We believe that the presented study is beneficial for the design of complex non-Hermitian systems in particular for sensing applications in optics and photonics. It allows to couple EPs to isolated eigenenergies but also to other EPs of variable order. Moreover, our theory can be used to search for high-response regions in a given exceptional surface. The divergence of the spectral response strength can be exploited as a numerical indicator for a higher-order EP when parameters are varied within the exceptional surface. This is analog to the usage of the Petermann factor~$\PF$ or the phase rigidity~$1/{\sqrt{\PF}}$~\cite{DMX16,Jin18,JBN23,XZH19,ZZJS20} as an indicator for EPs.

\acknowledgments 
Fruitful discussions with J. Kullig and H. Schomerus are acknowledged. 



%

\end{document}